\documentclass[%
 preprint,
 amsmath,amssymb,
 aps,floatfix
]{revtex4-2}

\usepackage{lipsum}
\usepackage{natbib}
\usepackage{url}
\usepackage[dvipsnames]{xcolor}
\usepackage{graphicx}
\usepackage{dcolumn}
\usepackage{bm}
\usepackage{siunitx}
\begin{document}

\preprint{}

\title{Probing optical spin-currents using THz spin-waves in noncollinear magnetic bilayers}

\author{Tom Lichtenberg}
  \email{t.lichtenberg@tue.nl}
\author{Maarten Beens}%
\author{Menno H. Jansen}%
\author{Bert Koopmans}%
\affiliation{
 Department of Applied Physics, Eindhoven University of Technology, P.O. Box 513, 5600 MB Eindhoven, Netherlands}%
\author{Rembert A. Duine}
\affiliation{
 Department of Applied Physics, Eindhoven University of Technology, P.O. Box 513, 5600 MB Eindhoven, Netherlands
and Institute for Theoretical Physics, Utrecht University, Leuvenlaan 4, 3584 CE Utrecht, Netherlands
}%
\date{\today}

\begin{abstract}
Optically induced spin currents have proven to be useful in spintronics applications, allowing for sub-ps all-optical control of magnetization. However, the mechanism responsible for their generation is still heavily debated. Here we use the excitation of spin-current induced THz spin-waves in noncollinear bilayer structures to directly study optical spin-currents in the time domain. We measure a significant laser-fluence dependence of the spin-wave phase, which can quantitatively be explained assuming the spin current is proportional to the time derivative of the magnetization. Measurements of the absolute spin-wave phase, supported by theoretical calculations and micromagnetic simulations, suggest that a simple ballistic transport picture is sufficient to properly explain spin transport in our experiments and that the damping-like optical STT dominates THz spin-wave generation. Our findings suggest laser-induced demagnetization and
spin-current generation share the same microscopic origin.
\end{abstract}

\maketitle

\begin{figure}[t!]
\includegraphics[width=10cm]{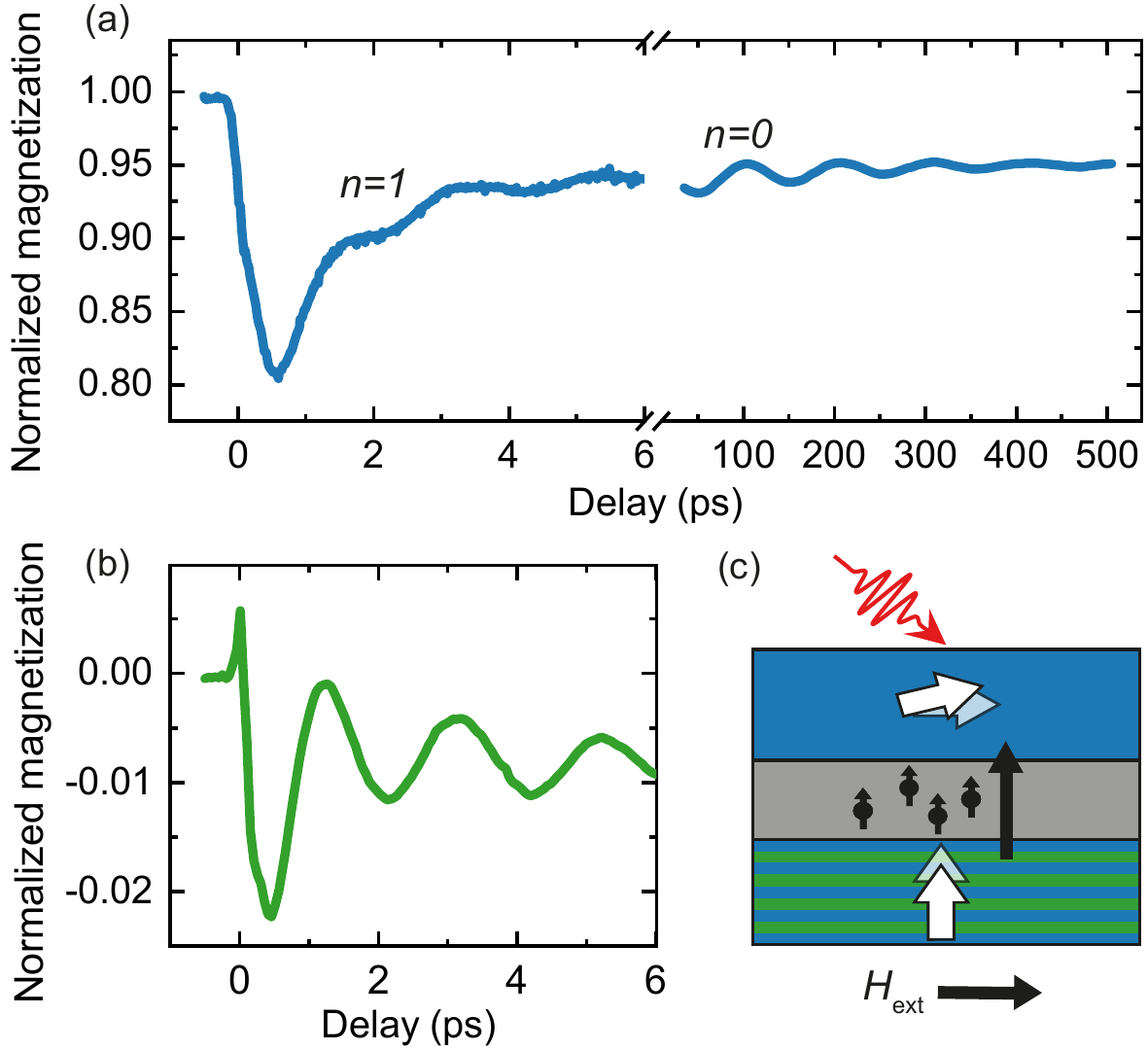}
\caption{\label{fig:1}(a) Magnetization dynamics of a noncollinear bilayer structure after fs laser pulse excitation. A laser fluence of $\approx$2.6 mJcm$^{-1}$ is used. The field is set to 0 and 91 mT respectively for the first-order inhomogeneous mode ($n=1$) and the homogeneous mode ($n=0$) measurements. (b) THz response of the absorption layer measured with Complex MOKE, normalized using the THz response in (a). (c) The sample geometry and OSTT mechanism.}
\end{figure}

{\it Introduction.}---Sub-ps demagnetization of ferromagnets (FM) upon fs laser-pulse excitation has garnered a lot of attention since its discovery in 1996 by Beaurepaire \emph{et al.} \cite{beaurepaire1996ultrafast}, both from a fundamental as well as a technological perspective. Over the past two decades, theoretical frameworks that describe this phenomenon in terms of local \cite{zhang2000laser, koopmans2005unifying, djordjevic2007connecting} and non-local \cite{Battiato2010SuperdiffusiveDemagnetization} angular momentum dissipation have been developed. Concurrently, its great potential in future data storage application was first demonstrated with the discovery of all-optical magnetization switching in ferrimagnetic alloys \cite{stanciu2007all}. \\
\indent The first experiments in collinear ferromagnetic bilayer structures showed that spin-angular momentum transfer between the two layers influenced both the speed and the magnitude of the laser-induced demagnetization of both layers \cite{Malinowski2008ControlMomentum}. In similar collinear systems, optical spin-currents have been used to enhance the efficiency and functionality of all-optical magnetization switching applications \cite{iihama2018single, VanHees2020, remy2020energy}. Studies of noncollinear bilayer systems have shown that optical spin-currents also allow for the control of the orientation of the magnetization \cite{Schellekens2014UltrafastExcitation, Choi2014SpinDemagnetization, choi2015thermal}. In these experiments, the polarization of the generated spin current is (nearly) perpendicular to the local magnetization, so the injected spin current exerts an optical spin-transfer torque (OSTT) on the magnetization. Recently, it was demonstrated that 90\% of this spin current is absorbed within $\sim$2 nm of the injection interface of the absorption layer, inducing a significant inhomogeneity in the magnetization. This results in the sub-ps all-optical excitation of highly tunable standing spin-waves (SSW) with frequencies in the THz regime \cite{Razdolski2017NanoscaleDynamics, Lalieu2017AbsorptionBilayers, lalieu2019investigating}.\\
\indent Optical spin-currents have been studied extensively in the last decade \cite{Battiato2010SuperdiffusiveDemagnetization, melnikov2011ultrafast, rudolf2012ultrafast, choi2014kerr, wieczorek2015separation, eschenlohr2017spin, hofherr2017speed, hurst2018spin, ulrichs2018micromagnetic, zhang2018laser, dewhurst2018laser, balavz2018transport}. However, the physics governing their generation and transport is still heavily debated. Several mechanisms have been proposed. A first idea is that upon laser excitation, mobile hot electrons are created in the FM. Battiato \emph{et al.} suggested that this hot electron current is polarized by the FM itself due to different lifetimes and velocities of excited majority and minority spins \cite{Battiato2010SuperdiffusiveDemagnetization, Battiato2012TheoryHeterostructures}. This results in a superdiffusive spin current flowing out of the FM, which contributes to the demagnetization process. A second possible mechanism, henceforth referred to as the $\text{d}M/\text{d}t$ mechanism, is based on electron-magnon coupling. Here, the demagnetization is due to magnon excitation, and the lost angular momentum is transferred to the mobile electrons in accordance with angular momentum conservation \cite{Choi2014SpinDemagnetization, choi2015thermal}. The resulting spin current is thus directly proportional to the the time derivative of the demagnetization. This implies laser-induced demagnetization and spin-current generation share the same microscopic origin. Recent theoretical work supports this view; the $s\text{-}d$ interaction, which mediates angular momentum transfer between local magnetic moments and itinerant electrons, is a promising candidate to explain this link between laser-induced demagnetization and spin-current generation \cite{tveten2015electron, shin2018ultrafast, beens}. Literature suggests the temporal profile of the generated spin currents scales differently with the laser-pulse energy depending on the excitation mechanism. Specifically, increasing the laser-pulse energy leads to an increase of demagnetization times and thus a broadening of the temporal profile of the spin current pulse if the $\text{d}M/\text{d}t$ mechanism is dominant \cite{koopmans2010explaining, kuiper2014spin}. Conversely, no such laser pulse energy dependence on the temporal profile of the spin current is expected in experiments where superdiffusive spin currents are dominant, as was previously argued in refs. \cite{vodungbo2012laser} and \cite{moisan2014investigating}.\\
\indent In this letter, we measure the phase of the THz spin-waves in noncollinear structures to directly study optical spin-currents in the time domain. The excited high-frequency dynamics allow us to study variations in the temporal profile of the spin current with a sub-picosecond resolution, which is sufficient to distinguish between the superdiffusive and $\text{d}M/\text{d}t$ mechanisms. We observe a significant phase shift of the THz spin-waves for increasing laser pulse energy, which strongly suggests the latter mechanism is dominant in our experiments. Micromagnetic simulations corroborate these findings. Furthermore, we argue based on our modelling and measurements of the absolute THz spin-wave phase that spin transport in our stacks is (close to) ballistic in nature. Lastly, we measured the phase of the optically excited homogeneous mode to show that the measured dynamics are driven by a pure damping-like torque. Our results demonstrate that the noncollinear bilayer magnetic structure is a powerful tool to not only study optical spin-current induced phenomena, but also the generation and short-scale transfer of optical spin-current themselves. \\   
\indent{\it Set-up and experiments.}---Our experiments are performed on a noncollinear bilayer structure consisting of  SiB(substrate)/{\allowbreak}Ta(4)/{\allowbreak}Pt(4){\allowbreak}/{\allowbreak}[Co(0.2)/{\allowbreak}Ni(0.6)]$_{4\times}$/{\allowbreak}Co(0.2){\allowbreak}/{\allowbreak}Cu(2.5){\allowbreak}/{\allowbreak}Co(5){\allowbreak}/{\allowbreak}Pt(2.5) (thickness in brackets is indicated in nm), as grown by magneton sputtering. The stack, including relevant excitation mechanisms, is illustrated in fig.\,\ref{fig:1}c. The Co/Ni multilayer is chosen for its PMA at relatively large magnetic volumes, making it an ideal thin film spin current generator \cite{kurt2010enhanced, johnson1996magnetic}. This out-of-plane (OOP) magnetized layer is separated from an in-plane (IP) magnetized Co layer with a thin Cu spacer, which facilitates spin transport between the two magnetic layers after laser excitation. Upon laser excitation, both magnetic layers demagnetize, generating an optical spin-current traveling into the other layer in the process. The polarization of this spin current is perpendicular to the local magnetization of the layer in which it is absorbed. As a consequence, the local magnetization obtains an OOP component due to the OSTT exerted by the spin current, thereby exciting dynamics in the absorption layer. In principle, both magnetic layers act as spin current generator and absorber. However, the sizeable PMA of the OOP layer suppresses the spin current induced deviation from equilibrium, so the amplitude of any excited spin waves is typically below the noise threshold
\cite{Lalieu2017AbsorptionBilayers}. Therefore, the OOP and IP layer will henceforth also be referred to as the generation and absorption layer respectively. \\
\indent The laser-induced dynamics are measured using pump-probe experiments, where the time-resolved magnetization is probed with the magneto-optical Kerr effect (MOKE) in the polar configuration. Laser pulses are generated by a Ti:sapphire laser at a repetition rate of 80 MHz and a wavelength of 780 nm. The pulse width at sample position is in the order of 150 fs. Both pump and probe pulses are focused onto the sample with a spot size of 16 and \SI{8}{\micro\metre} respectively. Experiments are done with laser fluences between 1.2 and 2.6 mJcm$^{-2}$. The so-called ``coherence peak", which arises due to pump-probe interference at temporal and spatial overlap, is used to properly define zero time delay \cite{eichler1984coherence, luo2009eliminate}. We refer to Sec.\,I of the supplementary information for a more elaborate explanation \cite{supplement}. A typical magnetization trace is shown in fig.\,\ref{fig:1}a. The OOP component of the magnetization of the complete stack is plotted as a function of the delay of the probe pulse after pump pulse excitation. This means the superposition of the demagnetization dynamics of the OOP layer and the optical spin current induced dynamics of the IP layer is visualized. On longer timescales, the spin current induced homogeneous mode ($n=0$) with frequencies in the order of 10 GHz can be observed, similar to the work discussed in ref. \cite{Schellekens2014UltrafastExcitation}. On a sub-10 ps timescale we observe the inhomogeneous higher-order standing spin-waves, of which the first order mode ($n=1$) with a frequency of $0.48\pm0.01$ THz is clearly visible \cite{Razdolski2017NanoscaleDynamics, Lalieu2017AbsorptionBilayers}. We employ complex MOKE to filter out the magnetic response of the generation layer to improve the signal to noise ratio in our measurements \cite{schellekens2014exploring}. It should be noted that significant leakage is unavoidable if the filtered signal is large compared to the dynamics of interest. The response of the absorption layer after the implementation of complex MOKE is shown in fig.\,\ref{fig:1}b. Because the contribution of the OOP dynamics to the MOKE signal is relatively large, it is impossible to filter it out completely. This leads to the complex behavior at small timescales and the general negative value of the measured signal.\\
\begin{figure*}[t!]
\includegraphics[width=17.2cm]{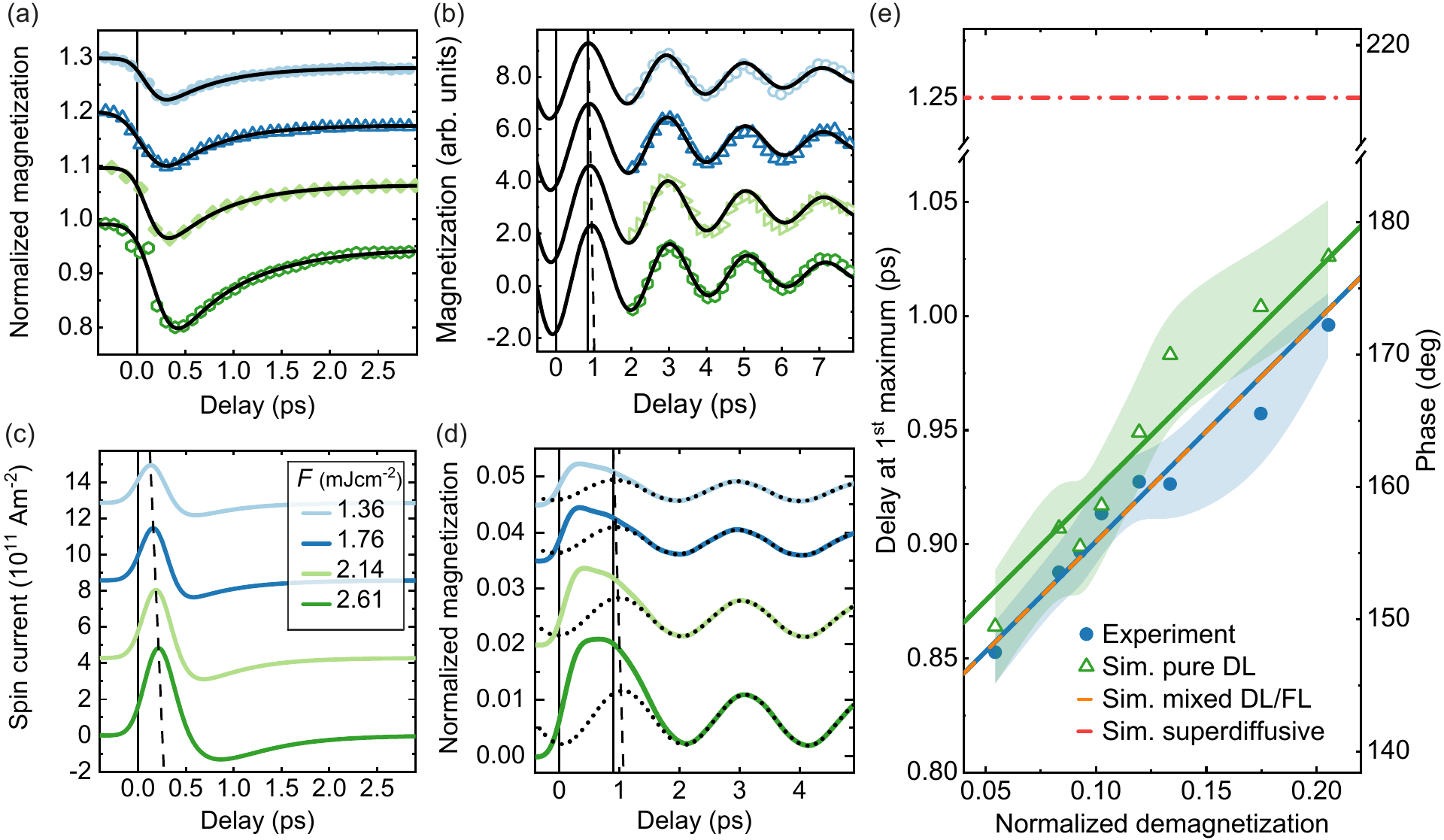}
\caption{\label{fig:2} (a) Demagnetization traces for 4 different laser fluences. The black lines are fits with the 3TM. (b) The measured THz spin-waves. The data is fitted with a damped cosine function, indicated with the black line. (c) The generated spin currents, calculated by taking the time derivative of the demagnetization traces. (d) Micromagnetic simulations of spin current induced dynamics in the absorption layer. The dotted line indicates fits with a damped sine function. The dashed lines provide a guide to the eye for the shift of the spin current maximum with respect to the vertical black line in (c) and the shift at the first maximum of the fit in (b) and (d) respectively. (e) The delay at the first maximum of the fit in (c) as a function of the normalized demagnetization extracted from the data in (a). The shaded areas and solid lines indicate the uncertainty regions and linear fits of the data with a shared slope. The dashed orange line indicates simulation results where a field-like contribution of 5.6\% to the spin current induced torque is assumed. The dash-dotted red line indicates the expected spin-wave phase using the spin current profile calculated in ref.\,\cite{ulrichs2018micromagnetic}. See Sec.\,V of the supplementary information for a full overview of the measurements and simulations.}
\end{figure*}
\indent{\it Spin-wave phase.}---In order to distinguish between the two aforementioned spin-current generation mechanisms, spin-current induced THz spin-waves are measured as a function of laser fluence, from which the experimental fluence-dependence of the spin-wave phase is extracted. These results are then compared to micromagnetic simulations where the expected fluence dependence of the spin-wave phase is determined numerically for both mechanisms. In fig.\,\ref{fig:2}a and b, the isolated fs laser-pulse induced magnetic response of the OOP and IP layer are plotted respectively for various laser fluences. The demagnetization of the OOP layer, as displayed in fig.\,\ref{fig:2}a, is fitted with an analytical solution of the phenomenological 3-Temperature model (3TM) \cite{dalla2007influence}. A clear enhancement of the demagnetization for increasing laser fluence can be observed, attributed to the increased heating of the sample. The spin-current induced THz spin-waves are shown for various laser fluences in fig.\,\ref{fig:2}b. Data for time delays below 2 ps are discarded because the spin current has not yet been absorbed completely and intermixing with the OOP demagnetization is significant. To extract the spin-wave phase, a damped cosine function is used to fit the data after a global background subtraction. The delay at the first maximum of this fit is used to define the spin-wave phase, as indicated by the dashed line. This delay is plotted as a function of the normalized maximum demagnetization of the OOP layer in fig.\,\ref{fig:2}e. A significant spin-wave phase shift as a function of demagnetization is observed, indicating that the changing demagnetization timescales significantly affect the temporal profile of the generated spin current. \\
\indent{\it Interpretation of experimental results.}---To further understand these results, we simulate the response of a one-dimensional IP magnetized layer, subdivided into 0.25 nm thick slabs, after excitation with an OOP polarized spin current. We employ the MUMAX$^3$ package \cite{vansteenkiste2014design} to solve the LLG equation including the Slonczewski spin-transfer torque term and calculate the magnetic response. Simulation- and material parameters are discussed in detail in Sec.\,II of the supplemental information \cite{supplement}. As previously mentioned, a laser-pulse induced superdiffusive hot-electron current is spin polarized by the ferromagnetic layer itself due to different lifetimes of excited majority and minority spins. The temporal profile of this spin current pulse depends only on the Gaussian temporal profile of the laser pulse and the thickness of the magnetic layer, and is independent of laser pulse energy \cite{Battiato2010SuperdiffusiveDemagnetization, ulrichs2018micromagnetic, ritzmann2020high, zhang2020ultrafast}. Thus, the THz spin-wave phase is expected to be constant in this scenario. The expected fluence dependence of the spin-wave phase based on calculations of the temporal spin-current profile as presented in ref.\,\cite{ulrichs2018micromagnetic} is drawn by the horizontal dot-dashed red line in fig.\,\ref{fig:2}e. However, it should be noted that the absolute value of this spin-wave phase depends on material and experimental parameters, and cannot be compared directly to the results presented in this work.\\
\indent In stark contrast to superdiffusive currents, spin currents generated with the $\text{d}M/\text{d}t$ mechanism are directly linked to the demagnetization process; spin-angular momentum lost during the demagnetization process is transferred to mobile electrons moving out of the ferromagnetic layer. Increasing the laser pulse energy leads to an increase of the timescales associated with the demagnetization \cite{koopmans2010explaining, moisan2014investigating, kuiper2014spin}. More specifically, this influences the temporal profile of the generated spin current, and thus the spin-wave phase. In order to quantify the expected fluence dependence of the spin-wave phase in this scenario, the change of the temporal profile of the generated spin current for increasing laser fluence is determined first. For this purpose, the time derivative of the fitted demagnetization traces of the OOP layers, as presented in fig.\,\ref{fig:2}a, is taken. To calculate the spin current absorbed by the absorption layer, spin transport through the Cu layer has to be taken into account. In Sec.\,IIIB of the supplementary information, a diffusive approach is discussed in detail \cite{supplement}. However, electron transport on length scales below the electron mean free path is not accurately captured by the diffusive model \cite{Battiato2010SuperdiffusiveDemagnetization, brorson1987femtosecond}, necessitating a different approach. Under the assumption that the absorption layer is an ideal spin sink for transverse spins and that spin transport between layers is ballistic (see Sec.\,IIIA of the supplementary material for the full derivation  \cite{supplement}), the spin current absorbed by the absorption layer (in units of Am$^{-2}$) can be written as 
\begin{align}
    J_{\text{S}} =-\varepsilon \frac{e\, d_\text{G} M_\text{S,G}}{\mu_{B}}\frac{\text{d}m_\text{G}}{\text{d}t}\, .
    \label{eq1}
\end{align}
Here, $d_\text{G} = 3.4$ nm, $M_\text{S,G} = 0.66$ MAm$^{-1}$ \cite{Lalieu2017AbsorptionBilayers} and $m_\text{G}$ are the thickness, saturation magnetization and normalized magnetization of the generation layer respectively. $e$ and $\mu_\text{B}$ are the electron charge and the Bohr magneton. Furthermore, $\varepsilon$ denotes the spin transfer efficiency between the two layers, which results from the relative importance of spin transport to the bottom Pt and local spin relaxation due to for instance Elliot-Yaffet processes \cite{koopmans2010explaining, supplement}. The spin current profiles are plotted in fig.\,\ref{fig:2}c, and used as the input of our micromagnetic simulations. \\
\indent The resulting magnetic response is depicted in fig.\,\ref{fig:2}d, with $\varepsilon\approx 0.06$ chosen such that the amplitude of the excited spin waves matches the experimental value. Our findings are in accordance with spin-transfer efficiency measurements done on our sample (see Sec.\,IV of the supplementary information), which yield $\varepsilon = 0.068 \pm 0.009$. Furthermore, similar values were found in these structures in the past (0.02-0.10) \cite{Schellekens2014UltrafastExcitation, Lalieu2017AbsorptionBilayers}. The sharp decrease of the amplitude after the first oscillation is attributed to the negative peak in the spin current, which corresponds to the remagnetization of the generation layer. For increasing laser fluence, the spin current amplitude increases and the temporal profile both broadens and shifts, leading to an increased spin-wave amplitude and phase respectively. Conversely, the spin-wave frequency is determined by the thickness of the absorption layer and does depend on laser fluence. Again, a damped cosine function is used to fit the THz response for time delays larger than 2 ps. The delay at the first maximum of the fit is plotted as a function of the total laser induced demagnetization in fig.\,\ref{fig:2}e. Here, the uncertainty is determined by both the 3TM fits in fig.\,\ref{fig:2}a and the damped cosine fit in fig.\ref{fig:2}d. A significant time shift of $\approx$150 fs is observed as a function of laser fluence, corresponding to a phase shift of about 30$^\circ$. This shift is attributed to the increase of the demagnetization times for increasing laser fluences, which leads to the temporal broadening of the spin current pulse, as well as to a shift in its maximum value, as indicated in fig.\,\ref{fig:2}c with the dashed line. Furthermore, the absolute phase of the spin-waves is 145-180$^\circ$. Intuitively, the spin wave phase is expected to coincide with the positive peak of the spin current, which is about 250 fs or 45$^\circ$ in our experiments. This discrepancy is attributed to the negative peak of the spin current hindering spin precession, leading to the change of the dynamics between 0 and 1 ps delay in fig.\,\ref{fig:2}b and thus an increase of the spin-wave phase. Within the error margin the measurements coincide with the simulations, which strongly suggests the $\text{d}M/\text{d}t$ mechanism is dominant in our experiments. Furthermore, the agreement between the measured and simulated absolute spin-wave phase suggests a relatively simple ballistic transport approach suffices to describe interlayer transport in these experiments. \\
\begin{figure}[t!]
\includegraphics[width=10cm]{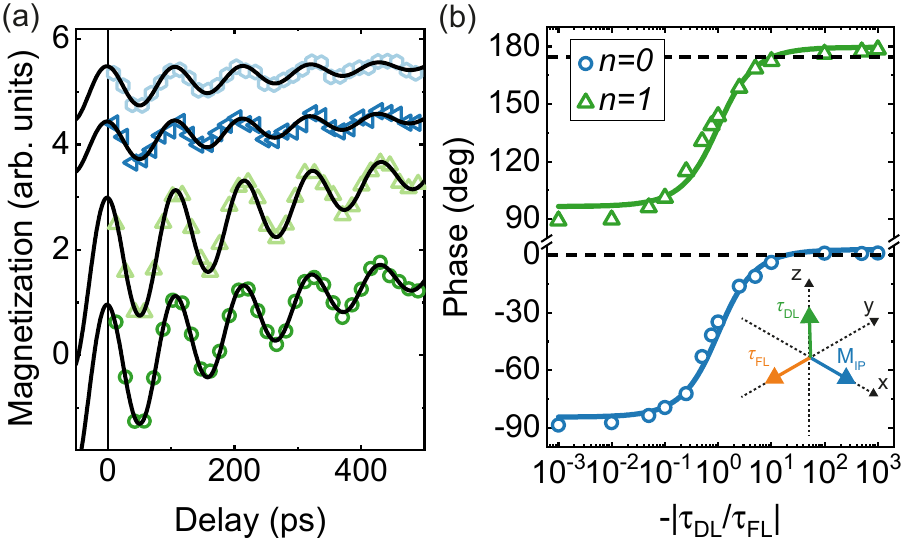}
\caption{\label{fig:3}(a) The homogeneous mode for $F=0.98 \text{ mJcm}^{-2}$ (blue datapoints) and $F=2.91 \text{ mJcm}^{-2}$ (green datapoints) in the examined fluence range at a magnetic field of 91 mT, for two magnetic field angles (light and dark datapoints for 0 and 3 degrees respectively) with respect to the magnetization of the absorption layer. (b) Simulated phase as a function of the ratio between FL and DL torque. The data is fitted with an arctangent and the dashed lines indicate experimental values. The inset indicates the direction of the absorption layer magnetization and the FL and DL torques.}
\end{figure}
\indent{\it Damping-like vs.\,field-like torques.}---Up until this point, the absorbed spin current is assumed to exert a pure DL torque due to the OSTT on the absorption layer. We use linearly polarized laser-light in our experiments, which means helicity-dependent Inverse spin-Hall effect induced field-like torques are not expected \cite{nvemec2012experimental, choi2017optical, choi2020optical}. Furthermore, our experiments are done at zero applied magnetic field and variations of the magnetization of the IP layer are small. Therefore, any significant contributions to a FL torque are expected to be small. However, phase measurements of the THz spin-waves allow us to study FL spin-transfer torque contributions in more detail and confirm these expectations. For this purpose, additional measurements of the homogeneous mode are done for two different laser fluences (fig.\,\ref{fig:3}a), which are again fitted with a damped cosine function. For the entire studied fluence range, the first maximum of the fit corresponds to zero time delay within the margin of error, corresponding to a spin-wave phase shift of $(0\pm 1)^\circ $. This result again highlights the necessity of using high-frequency spin waves to study spin currents with sufficient time resolution, as the error margin is too large to observe the sub-ps shifts shown in fig.\,\ref{fig:2}e. These results are further validated by comparing measurements at a magnetic field angle of $\phi=0^\circ$ and $\phi=3^\circ$ with respect to the IP direction; small variations of the applied magnetic field do not induce a significant phase shift. MUMAX$^3$ has been employed to simulate the phase of both the homogeneous mode and the 1$^\text{st}$ order mode as a function of the ratio between the FL and DL spin-transfer torque \cite{vansteenkiste2014design}. Refer to Sec.\,VI of the supplemental information for the simulation details. The results are plotted in fig.\,\ref{fig:3}b. Because the FL torque points in the IP direction and the DL torque in the OOP direction, as sketched in the inset of fig.\,\ref{fig:3}b, a 90 degree phase shift is expected when the FL torque becomes dominant, and the behaviour can in general be described by an arctangent function. From these simulations it can be concluded that the DL OSTT is dominant in our experiments, which is in line with previous theoretical considerations of spin current induced THz spin-waves \cite{ulrichs2018micromagnetic, ritzmann2020high} and experiments using linearly polarized laser light \cite{nvemec2012experimental, choi2017optical, choi2020optical}. Assuming a negative FL contribution of about 5.6\%, the simulated and measured THz spin-waves completely overlap, as indicated in fig. \ref{fig:2}e. The ratio between the FL and DL torque is given by the ratio between the imaginary and real part of the spin-mixing conductance at the Cu/Co interfaces, which is typically small for a nonmagnetic/transition metal interface \cite{charilaou2010spin, tserkovnyak2005nonlocal}. Our results are reasonably in line with first principle calculations of this ratio for Cu(1 1 1)/Co, which is in the order of 2.2-2.8 \% \cite{xia2002spin, zwierzycki2005first}. However, it should be noted that the uncertainty of the THz spin-wave phase measurements is relatively large compared to the phase variation in fig.\,\ref{fig:3}b. Helicity dependent experiments can be done in the future to study FL-torque contributions to the optical spin-current induced dynamics, and in particular THz spin-waves, in noncollinear structures in more detail. \\
\indent{\it Conclusion.}---We have shown that noncollinear bilayer structures are ideal tools to study the rich physics of fs laser-pulse induced magnetization dynamics in magnetic multilayers. We measured the optical spin-current induced THz spin-wave phase as a function of laser fluence to study the temporal profile of the optical spin-current itself. We find that the THz spin-wave phase increases with increasing laser fluence, strongly indicating the spin current is proportional to the time derivative of the laser-induced demagnetization, which means that optical spin-currents and ultrafast demagnetization are driven by the same mechanism in our experiments. We further demonstrated that only considering ballistic interlayer-transport is enough to fully explain our observations. Lastly, it has been shown that the DL OSTT-mechanism dominates spin transfer from the spin current to the local magnetization of the absorption layer.\\
\indent This work is part of the research program of the Foundation for Fundamental Research on Matter (FOM), which is part of the Netherlands Organization for Scientific Research (NWO).

\newpage
\bibliography{Main}

\end{document}


\preprint{}

\title{Supplemental material: Probing optical spin-currents using THz spin-waves in noncollinear magnetic bilayers}

\author{Tom Lichtenberg}
  \email{t.lichtenberg@tue.nl}
\author{Maarten Beens}%
\author{Menno H. Jansen}%
\author{Bert Koopmans}%
\affiliation{Department of Applied Physics, Eindhoven University of Technology, P.O. Box 513, 5600 MB Eindhoven, Netherlands}

\author{Rembert A. Duine}
\affiliation{
 Department of Applied Physics, Eindhoven University of Technology, P.O. Box 513, 5600 MB Eindhoven, Netherlands
and Institute for Theoretical Physics, Utrecht University, Leuvenlaan 4, 3584 CE Utrecht, Netherlands
}%
\date{\today}

\maketitle
\beginsupplement
\newpage
\section{Extraction of the absolute spin wave phase} \label{SEC:SE1}
\begin{figure}[h!]
\includegraphics[width=8cm]{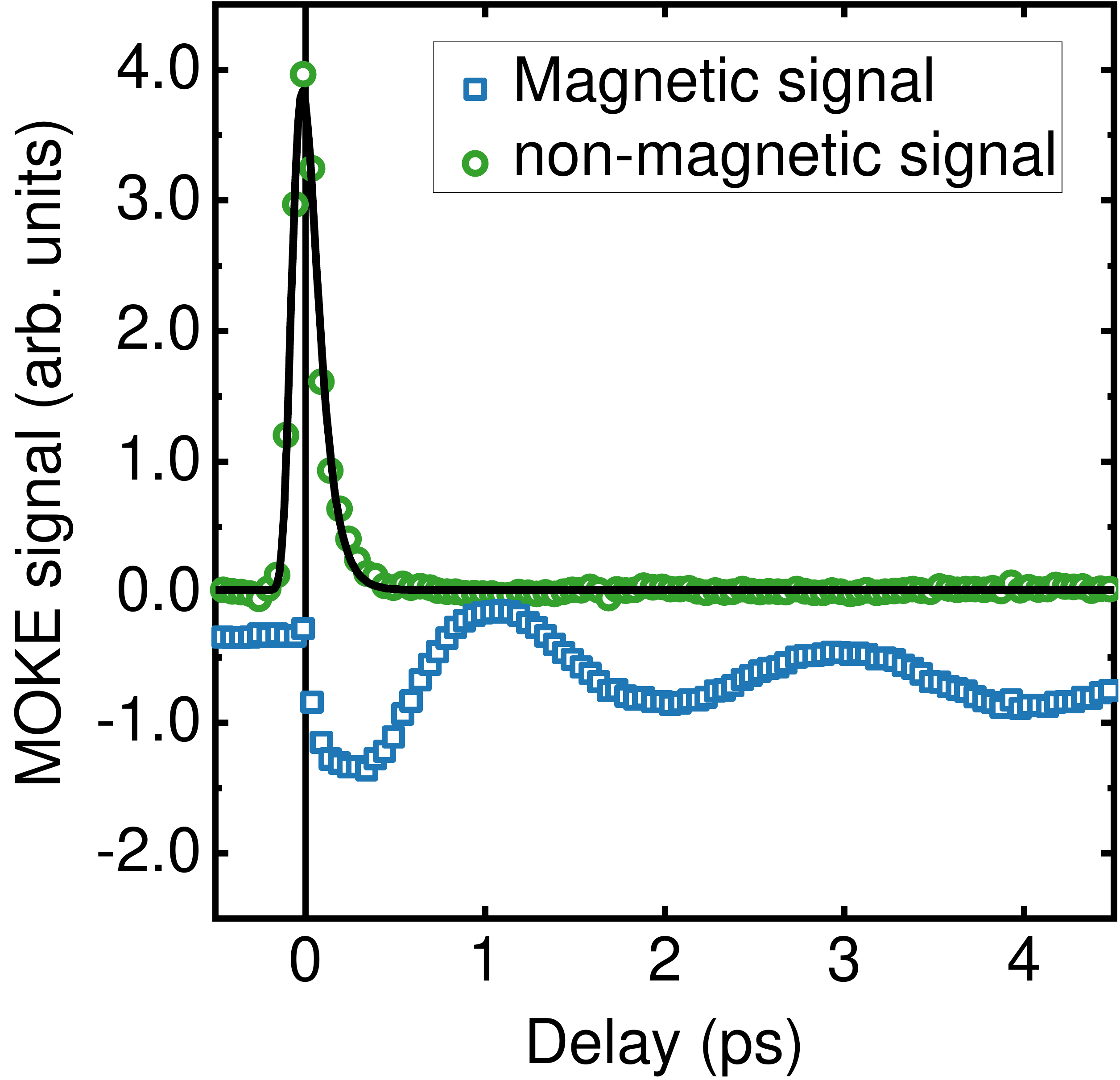}
\caption{\label{fig:S1}Difference and sum of the spinwave measurements at positive and negative field, yielding the magnetic and non-magnetic part of the measured signal. The black line indicates a peak fit using the \textit{Origin} software package.}
\end{figure}

\noindent To determine the absolute phase of the measured spin waves, the moment at which pump and probe overlap in time is determined. For this purpose, additional measurements are done at high time resolution and both negative and positive applied magnetic field. This allows us to extract both the magnetic and non-magnetic parts of the signal by taking the difference and sum of the measured dynamics respectively. The results, plotted in fig.\,\ref{fig:S1}, show a significant non-magnetic contribution, which is attributed to coherence effects during pump-probe overlap \cite{eichler1984coherence, luo2009eliminate}. The data are fitted using the peak fitting tool in the \textit{Origin} software package, from which the temporal pump-probe overlap is determined. The determined uncertainty in this value is added to the uncertainty of the damped cosine fits in fig.\,2b of the main text.
\newpage
\section{Micromagnetic simulation parameters} \label{SEC:SE2}
\begin{figure}[h!]
\includegraphics[width=10cm]{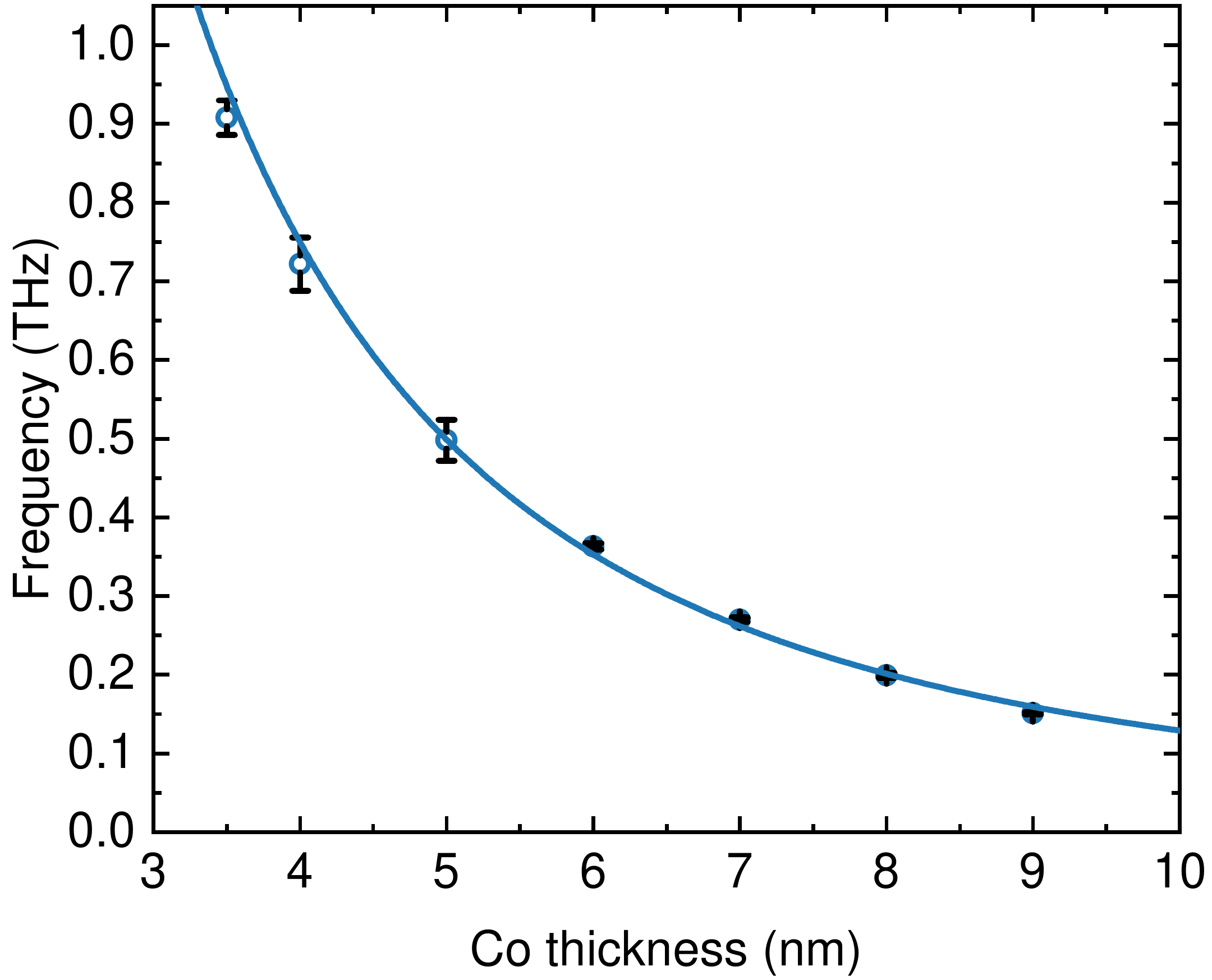}
\caption{\label{fig:S2} Measured spin wave frequency as a function of the Co absorption layer thickness, fitted with spin wave dispersion function eq. S1.}
\end{figure}

\noindent For our simulations, a $25 \times 25 \times 5$ nm$^3$ large IP magnetized Co layer is defined, subdivided into $N_\text{x} \times N_\text{y} \times N_\text{z} = 100 \times 100 \times 20$ cells. To simulate an infinite layer, it is surrounded by 5 copies in the $x$ and $y$ direction. This is further tested by comparing results at the edges and the center of the layer, which are identical. The used material parameters are $M_\text{S}=1.24$ MAm$^{-1}$, $K_\text{S}=0.7$ mJm$^{-2}$ \cite{Lalieu2017AbsorptionBilayers}. The exchange parameter $A_\text{ex}$ is determined by measuring the spin wave dispersion. In fig.\,\ref{fig:S2}, the spin wave frequency is plotted as a function of the thickness of the Co absorption layer.  
\noindent The data can be fitted with a dispersion relation for standing exchange spin waves, as derived in ref.\,\cite{Lalieu2017AbsorptionBilayers}:
\begin{equation}
\begin{array}{c}
f(k)=\frac{\gamma}{2 \pi} \sqrt{\left(B_{\mathrm{app}}+\frac{2 A_{\mathrm{ex}}}{M_{\mathrm{s}}} k^{2}\right)\left(B_{\mathrm{app}}+\mu_{0} M_{\mathrm{s}}-\frac{2 K_{\mathrm{s}}}{t M_{\mathrm{s}}}+\frac{2 A_{\mathrm{ex}}}{M_{\mathrm{s}}} k^{2}\right)}\,, \\
k=\frac{\pi n}{t}\,.
\end{array}
\end{equation}
\noindent Here, $B_\text{app}$ is the applied field, $\gamma$ is the gyromagnetic ratio, $\mu_0$ the Bohr magneton, $A_\text{ex}$ the exchange parameter, $M_\text{S}$ the saturation magnetization, $K_\text{S}$ the surface anisotropy and $n$ the mode number of the excited spin wave. The measurements are done without an applied magnetic field and we are only considering the first order mode ($n=1$). Using the parameters given in ref.\,\cite{Lalieu2017AbsorptionBilayers}, we can extract $A_\text{ex}=30\pm 2$ pJm$^{-1}$, which is 1.5 - 2.0 times larger than typically measured in fcc Co \cite{vaz2008magnetism}. However, this disparity is beyond the scope of this letter, and we use the fitted exchange parameter for all simulations presented in this work in order to simulate spin waves with the correct dispersion. See refs.\,\cite{ritzmann2020high} and \cite{lalieu2017absorption} for further comments on this issue.\\
\indent Furthermore, a damping coefficient of 0.01 is assumed. The optical spin current is injected from the bottom and is polarized in the $z$-direction. The spin current is assumed to exert a pure damping-like (DL) torque on the magnetization \cite{ulrichs2018micromagnetic, ritzmann2020high}. The spin current absorption length is 0.96 nm, in accordance with ref. \cite{Lalieu2017AbsorptionBilayers}. In our experiments, we are mostly sensitive to the topmost part of the absorption layer due to laser attenuation. To account for this effect, exponential laser attenuation is assumed with a typical length scale of 13 nm \cite{johnson1974optical}. The magnetic response of each single-celled layer is weighted with this exponential function. Any enhanced MOKE effects at the interfaces are neglected. \newpage

\section{Interlayer spin transport}
\subsection{Derivation of equation 1}

\noindent In this section we present the derivation of equation 1 from  the main text. We define a spin current $J_s$ (in units $\mbox{Am}^{-2}$), that describes the net (OOP polarized) spin flow from the OOP generation layer to the IP absorption layer. The longitudinal spin dynamics within the generation layer is described in terms of the spin accumulation $\mu_\text{s}$ \cite{Choi2014,Tveten2015,Kimling2017}  

\begin{eqnarray}
\label{eq:s1}
\dfrac{d\mu_{s,G}}{dt} 
&=& 
\dfrac{1}{N_F}
\dfrac{M_{s,G}}{\mu_B}  \dfrac{dm_G}{dt} 
- \dfrac{\mu_{s,G}}{\tau_{s,G}}
-\dfrac{J_s}{ (-e)d_G N_F},
\end{eqnarray}

\noindent where the first term on the right-hand side corresponds to the source term arising from the local $s$-$d$ interaction \cite{Tveten2015,Choi2014,Kimling2017,Shin2018}, with $N_F$ the spin-averaged density of states at the Fermi level, $M_{s,G}$ the saturation magnetization of the generation layer, $m$ is the normalized magnetization, $e$ is the electron charge and $\mu_B$ the Bohr magneton. The second term describes the spin-flip processes, parametrized by time scale $\tau_{s,G}$. The third term expresses the spins pumped out of the generation layer. The factor $d_G^{-1}$, with $d_G$ the generation layer thickness, arises from the area to volume ratio. 

We assume transport in the spacer (Cu) layer is ballistic. Then, the exchange of spin between the generation layer and absorption layer is determined by the difference in spin accumulation of the layers. This approach is similar to Section IV of ref.\,\cite{beens2020s}. Since the IP layer absorbs the spins with an OOP polarization very efficiently, we assume that the absorption layer acts as an ideal spin sink and set the spin accumulation (specifically, its OOP component) to zero $\mu_{s,A}=0$. This yields the following expression for the interlayer spin current 

\begin{eqnarray}
\label{eq:s2}
J_s &=&(-e) \dfrac{G}{\hbar} \mu_{s,G} ,
\end{eqnarray}

\noindent where $G$ is an effective conductance expressed in units $\mbox{m}^{-2}$ and includes the electronic and spin-mixing conductance of both interfaces.  Now we substitute this expression in  eq.\,\ref{eq:s1} and simplify

\begin{eqnarray}
\label{eq:s3}
\dfrac{d\mu_{s,G}}{dt} &=& 
\dfrac{1}{N_F} \dfrac{M_{s,G}}{\mu_B} 
\dfrac{dm_G}{dt} 
-\dfrac{\mu_{s,G}}{\tau} ,
\end{eqnarray}

\noindent where we introduced the time scale $\tau^{-1} = \tau_{s,G}^{-1} +\tau_B^{-1} $ with definition $\tau_B^{-1}=G/(\hbar d_G N_F)$. The time scale $\tau$ functions as a response time. In the limit that $\tau$ is the shortest time scale in the system, which can correspond to either an ultrashort $\tau_B$ or $\tau_{s,G}$, the temporal profile of the spin accumulation will be directly determined by the source term. In this limit the spin accumulation is given by 

\begin{eqnarray}
\label{eq:s4}
\mu_{s,G} &=& \tau \dfrac{1}{N_F} 
\dfrac{M_{s,G}}{\mu_B} 
\dfrac{dm_G}{dt} .
\end{eqnarray}

\noindent Substituting this back into eq.\,\ref{eq:s2} yields 

\begin{eqnarray}
\label{eq:s5}
J_s &=& (-e) \dfrac{1}{\dfrac{\tau_B}{\tau_{s,G} }+1} 
\dfrac{d_G M_{s,G}}{\mu_B} \dfrac{dm_G}{dt}.
\end{eqnarray}

\noindent Note that the prefactor describes the ratio between the local spin loss (by $\tau_{s,G}$) compared to the spins transported towards the absorption layer. To account for additional spin leakage, e.g., towards the neighbouring Pt layer or spin-flip processes at the interfaces, we replace the prefactor by a phenomenological efficiency parameter $\varepsilon$. Then, the spin current is expressed as

\begin{eqnarray}
\label{eq:s6}
J_s &=& \varepsilon
\dfrac{ (-e) d_G M_{s,G}}{\mu_B} 
\dfrac{dm_G}{dt} ,
\end{eqnarray}

\noindent which is equation 1 of the main text. This analytical approximation is valid as long as the response $\tau$ is relatively short. Explicit calculations are presented in fig.\,\ref{fig:fig1}, where we compare the spin current that follows from directly solving eq.\,\ref{eq:s1} to the analytical approximation in Eqs.\,\ref{eq:s5}) and\,\ref{eq:s6}. We use the example source given by \cite{Choi2014}

\begin{eqnarray}
\label{eq:s7}
\dfrac{dm_G}{dt} &=& [A_1\cdot \exp(-(t-t_1)^2/t_2)
+A_2\cdot \exp(-(t-t_3)^2/t_4)] ,
\end{eqnarray}

\noindent that describes the standard bipolar behavior. Here, $A_1=0.07$ ps$^{-1}$, $A_2=-0.23$ ps$^{-1}$, $t_1=1.9$ s, $t_2=1.0$ s, $t_3=1.0$ s and $t_4=0.2$ s.\\
\indent Fig.\,\ref{fig:fig1}a presents the calculations for the material parameters that correspond to Ni (for the Co/Ni multilayer), summarized in table\,\ref{tab:tab1}. We included a prefactor arranging the source term in units $\mbox{Am}^{-2}$ (eq.\,\ref{eq:s6} with $\varepsilon=1.0$). We find response time $\tau\sim 75\mbox{ fs}$ (which leads to an additional 75 fs shift in the experiments presented in fig.\,2e in the main text) and efficiency $\varepsilon\sim 0.24$ (estimated via the prefactor in eq.\,\ref{eq:s5}). The calculations do not fully explain the experimental results, where no significant shift is observed, as well as a lower spin transfer efficiency. We argue that this discrepancy arises because we did not include the leakage of spins towards the Pt layer, and we do expect that the presence of multiple interfaces (as we have a Co/Ni multilayer and a Co/Pt interface) will effectively enhance the spin-flip relaxation rate. This can phenomenologically be introduced by using a shorter spin-flip relaxation time. For example, if we take the spin-flip time scale of a Co/Pt multilayer ($\tau_{s,G}=0.02\mbox{ ps}$), an efficiency of $\varepsilon=0.06$ is found, which is close to the experimental value. Furthermore, the shift is reduced to about 19 fs, which is within the margin of error in the experiments. Fig.\,\ref{fig:fig1}b shows the result for this enhanced spin-flip rate, suggesting a good correspondence between eq.\,\ref{eq:s6} (limit $\tau\rightarrow 0$) and the complete solution. This supports the validity of an approach as presented in eq.\,1 of the main text and we conclude that a simple ballistic description suffices to explain the experimental implications. However, we stress that eq.\,\ref{eq:s6} remains an approximation to the complete result, since the response time $\tau$ will always be finite. 
\subsection{Ballistic transport versus spin diffusion} 

\begin{figure}[t!]
\includegraphics[scale=0.9]{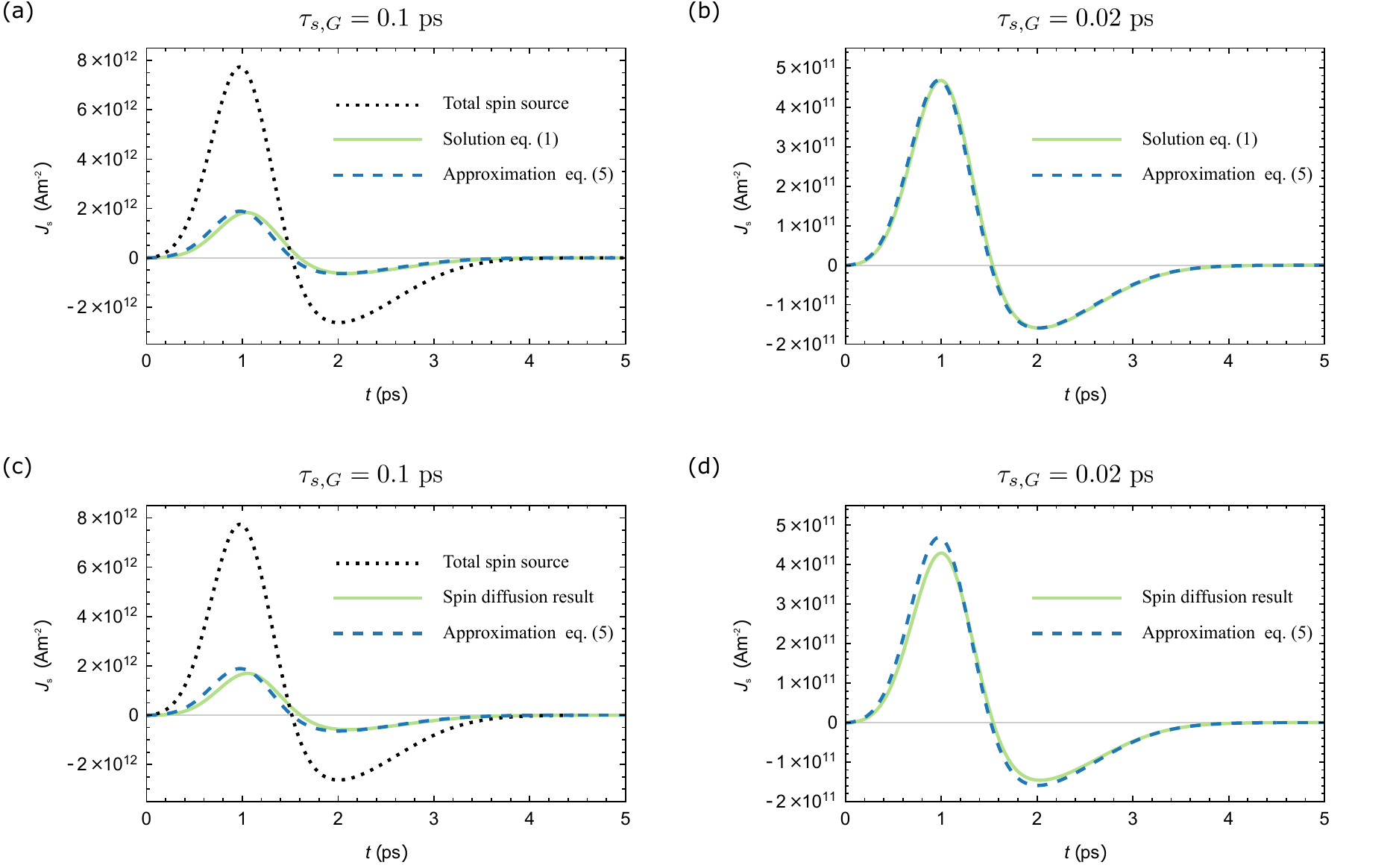}
\caption{\label{fig:fig1} Calculations of the interlayer spin current using a ballistic (a)-(b) and diffusive (c)-(d) description. (a) The ballistic calculation for the material constants of Ni. The black dotted curve indicates the total spin source per unit area (eq.\,\ref{eq:s6} with $\varepsilon = 1.0$). The solid green curve indicates the numerical result (solving eq.\,\ref{eq:s1}). The dashed blue curve indicates the analytical approximation, as introduced in eq.\,\ref{eq:s6}. (b) shows the ballistic calculation using an enhanced spin-flip scattering time of $\tau_{s,G}=0.02\mbox{ ps}$. (c) The diffusive calculation for Ni, indicated by the solid  green curve. (d) The diffusive calculation for $\tau_{s,G}=0.02\mbox{ ps}$. Note that the subfigures have different scalings of the $y$ axis. } 
\end{figure}

\noindent A similar result can be derived by solving the spin diffusion equation. The results are shown in fig.\,\ref{fig:fig1}c-d. Although the simulation has similar results compared to the ballistic calculation in last section, we note that the validity of this diffusive approach is questionable for such small layer thicknesses. We here present a brief discussion of the diffusive approach, for details we refer to ref.\,\cite{beens2020s}. 

We define the OOP spatial coordinate $x$ in the domain $x\in[-d_G,d_{\mathrm{Cu}}]$. The interface between the generation layer and spacer layer is located at $x=0$. The second interface is located at $x=d_\mathrm{Cu}$. Beyond the second interface the spin accumulation is set to zero to simulate the absorption layer acting as an ideal spin sink. The spin current at $x=-d_\mathrm{G}$ is set to zero. The continuity equation for the spins can be expressed as \cite{beens2020s, Kimling2017}  

\begin{eqnarray}
\label{eq:s8}
\dfrac{\partial \mu_s}{\partial t} &=& 
\dfrac{1}{N_F(x)}\dfrac{M_{s,G}}{\mu_B} 
\dfrac{dm_G}{dt} +
\dfrac{1}{N_F(x)} 
\dfrac{\partial }{\partial x}
\bigg(\dfrac{\sigma(x)}{e^2} 
\dfrac{\partial \mu_s}{\partial x} 
\bigg)-
\dfrac{\mu_s}{\tau_s(x)} ,
\end{eqnarray} 

\noindent where $\sigma(x)$ indicates the spin-averaged electrical conductivity and is a function of the spatial coordinate $x$. The source term is only nonzero in the ferromagnetic region. The interfacial spin current between the generation layer and spacer layer is expressed as 

\begin{eqnarray}
\label{eq:s9}
J^{\mathrm{int}}_{s,\mathrm{Co/Cu}} &=& 
(-e) \dfrac{G_{\mathrm{Co/Cu}}}{\hbar} 
(\mu_s(0^-)-\mu_s(0^+))  ,
\end{eqnarray}

\noindent whereas the spin current into the sink (absorption layer) is given by 

\begin{eqnarray}
\label{eq:s10}
J^{\mathrm{sink}}_s &=& 
(-e) \dfrac{G_{\mathrm{Co/Cu}}}{\hbar} 
\mu_s(d_\mathrm{Cu}^-).
\end{eqnarray}

\noindent The spin current $J_s^{\mathrm{sink}}$ resulting from the spin source (Eq.\,\ref{eq:s7}) is plotted in Figs.\,\ref{fig:fig1}(c)-(d), for the same system parameters as the ballistic calculation. The figures show that the two calculations give similar results. This can be understood from the fact that the long spin diffusion length of the Cu layer leads to an approximately spatially constant spin current. Analogously to the ballistic calculation, changing the spin-flip relaxation time to the value for Co/Pt yields an efficiency that is close to the experimental value, as shown in Fig. \ref{fig:fig1}(d). 

We show that both approaches yield similar results when applied to the noncollinear system studied in the main text. However, since a diffusive model is not suitable for electron transport on length scales below the mean free path it is not completely valid. We conclude this supplemental section by stressing that the ballistic assumption and its resulting analytical approximation (eq.\,1 in the main text) suffices to explain the experimental result.

\begin{table}[ht]
\centering
\caption{\label{tab:tab1} Numerical values used in the calculations.}
\begin{tabular}[t]{lccc}
\hline
symbol &  meaning & estimate & unit \\
\hline
$\mu_B$ & Bohr magneton & $9.27\times 10^{-24}$ & $\mbox{Am}^2$ \\
$M_{s,G}$ & Saturation magnetization generation layer \footnotemark[1]& $ 0.66\times 10^6 $ & $\mbox{Am}^{-1} $ \\ 
$d_G$ & Generation layer thickness \footnotemark[1] & $3.4$ & \mbox{ nm}\\
$d_{\mathrm{Cu}}$ & Spacer layer thickness \footnotemark[1] & $2.5$ & $\mbox{nm}$ \\ 
$\tau_{s,\mathrm{Co/Ni}}$ & spin relaxation time Co/Ni  \footnotemark[2]& $ 0.1 $ & $\mbox{ps}$ \\ 
$\tau_{s,\mathrm{Co/Pt}}$ & spin relaxation time Co/Pt \footnotemark[2] & $0.02$ & $\mbox{ps}$ \\
$\sigma_\mathrm{Ni}$ & electrical conductivity Ni \footnotemark[2] & $7.1\times 10^6$ & $\mbox{Sm}^{-1} $\\ 
$\sigma_\mathrm{Cu}$ & electrical conductivity Cu \footnotemark[2] & $39\times 10^6$ & $\mbox{Sm}^{-1} $\\ 
$D_\mathrm{Ni}$ & diffusion coefficient Ni \footnotemark[2] & $160$ & $\mbox{nm}^{2}\mbox{ps}^{-1}  $\\ 
$D_\mathrm{Cu}$ & diffusion coefficient Cu \footnotemark[2] & $9500$ & $\mbox{nm}^{2}\mbox{ps}^{-1}  $\\ 
$N_{F\mathrm{,Ni}}$ & density of states Ni \footnotemark[3] & $277$ & $\mbox{eV}^{-1}\mbox{nm}^{-3} $\\ 
$N_{F\mathrm{,Cu}}$ & density of states Cu \footnotemark[3] & $26$ &$\mbox{eV}^{-1}\mbox{nm}^{-3} $\\ 
$G$ & effective interface spin conductance \footnotemark[4] & $0.2\times 10^{19} $ & $\mbox{m}^{-2}$ \\ 
\hline
\footnotetext{Given in main text} 
\footnotetext{Taken from \cite{Shin2018}}
\footnotetext{Calculated from the relation $D=\sigma/(e^2 N_F)$}
\footnotetext{Estimated via its relation to the interfacial electrical conductance $G=\hbar G_e/(2e^2)$ using the interfacial conductance $G_e$ for a Co/Cu interface of $G_e = 2\times 10^{15} \mbox{ Sm}^{-2}$\cite{Shin2018}. For a single interface this yields, $G_{\mathrm{Co/Cu}}=0.4\times 10^{19}\mbox{ m}^{-2}$. We estimate the effective $G$ as two identical interfaces in series, which leads to an extra factor of $1/2$.  } 
\end{tabular}
\end{table}%

\clearpage
\newpage
\section{Spin transfer efficiency}
\begin{figure}[h!]
\includegraphics[scale=1]{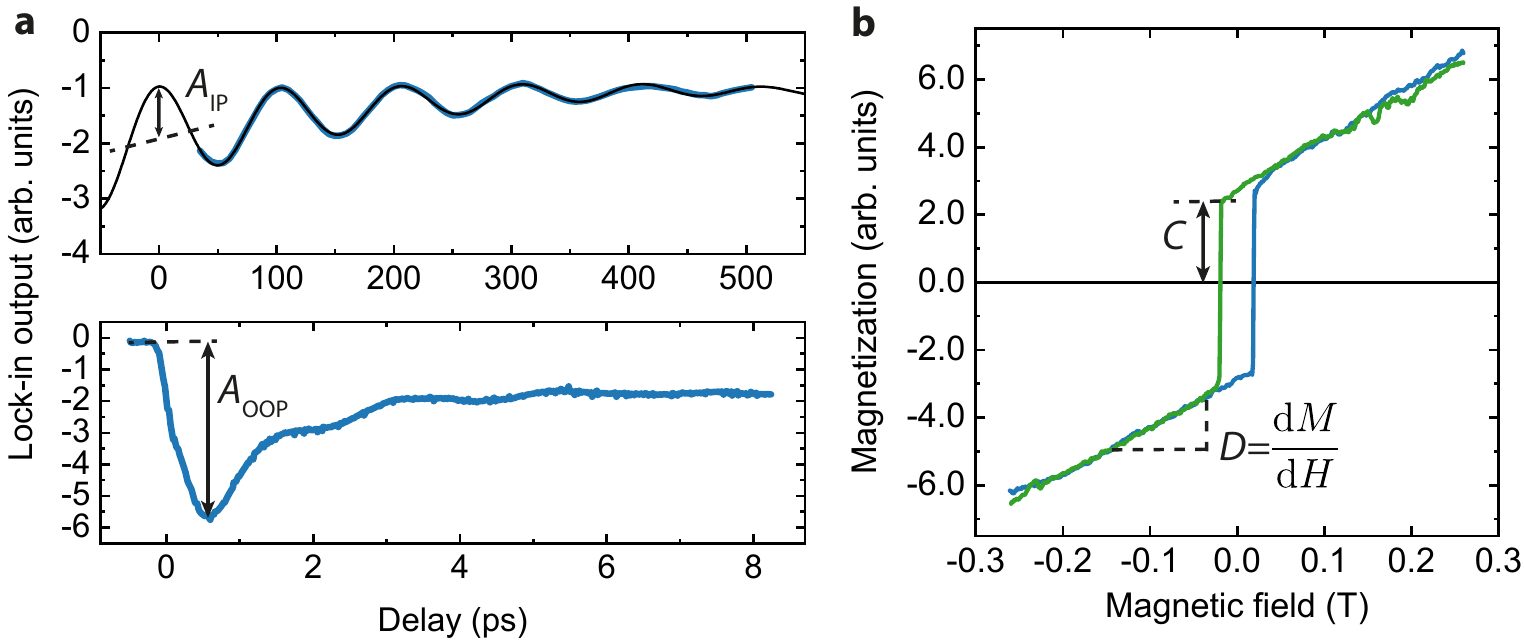}
\caption{\label{fig:S7} a) Lock-in output for the homogeneous mode and demagnetization measurements. The homogeneous mode is fitted with a damped cosine function, from which $A_\text{IP}$ can be extracted b) Hysteresis loop in the polar MOKE configuration, after extraction of Faraday effect. The relevant parameters are indicated in the figure.} 
\end{figure}

\noindent The interlayer spin transfer efficiency can be measured directly by determining the ratio between the angular momentum absorbed by the absorption layer and the angular momentum lost by the generation layer during demagnetization ($A_\text{IP}/A_\text{OOP}$). This ratio can be extracted directly from fig.\,\ref{fig:S7}a, where the output signal of our experiment is plotted.  
\noindent The relative optical sensitivity to both layers $f_\text{MO}$ has to be taken into account due to the optical nature of our experiments. The spin transfer efficiency is thus given by \cite{lalieu2017absorption} 
\begin{align}
\label{eq:sX}
\varepsilon &= f_\text{MO} \frac{A_\text{IP}}{A_\text{OOP}}\,,\\
f_\text{MO} &= \frac{C}{D}\frac{t_\text{A} {M_\text{S,A}}^2}{t_\text{G}| \frac{2 K_\text{S}}{t_\text{A}}-\mu_0 {M_\text{S,A}}^2| M_\text{S,G}}\,.
\end{align}
\noindent Here, $C$ and $D$ are extracted from the hysteresis loop shown in fig.\,\ref{fig:S7}b. The measurements are done in the polar MOKE configuration, so $C$ and $D$ correspond to the stepsize of the switch of the OOP layer and the slope of the hard axis rotation of the IP layer respectively. The latter value is corrected for the Faraday effect in optical components during the measurement. Solving the equation using parameters given in the main text as well as section II, an efficiency of $\varepsilon = 0.068 \pm 0.009$ is found. This value corresponds well to the simulations presented in the main text, as well as measurements in similar stacks \cite{lalieu2017absorption, Schellekens2014UltrafastExcitation}.
\newpage
\section{Complete datasets}
\begin{figure}[h!]
\includegraphics[width=12cm]{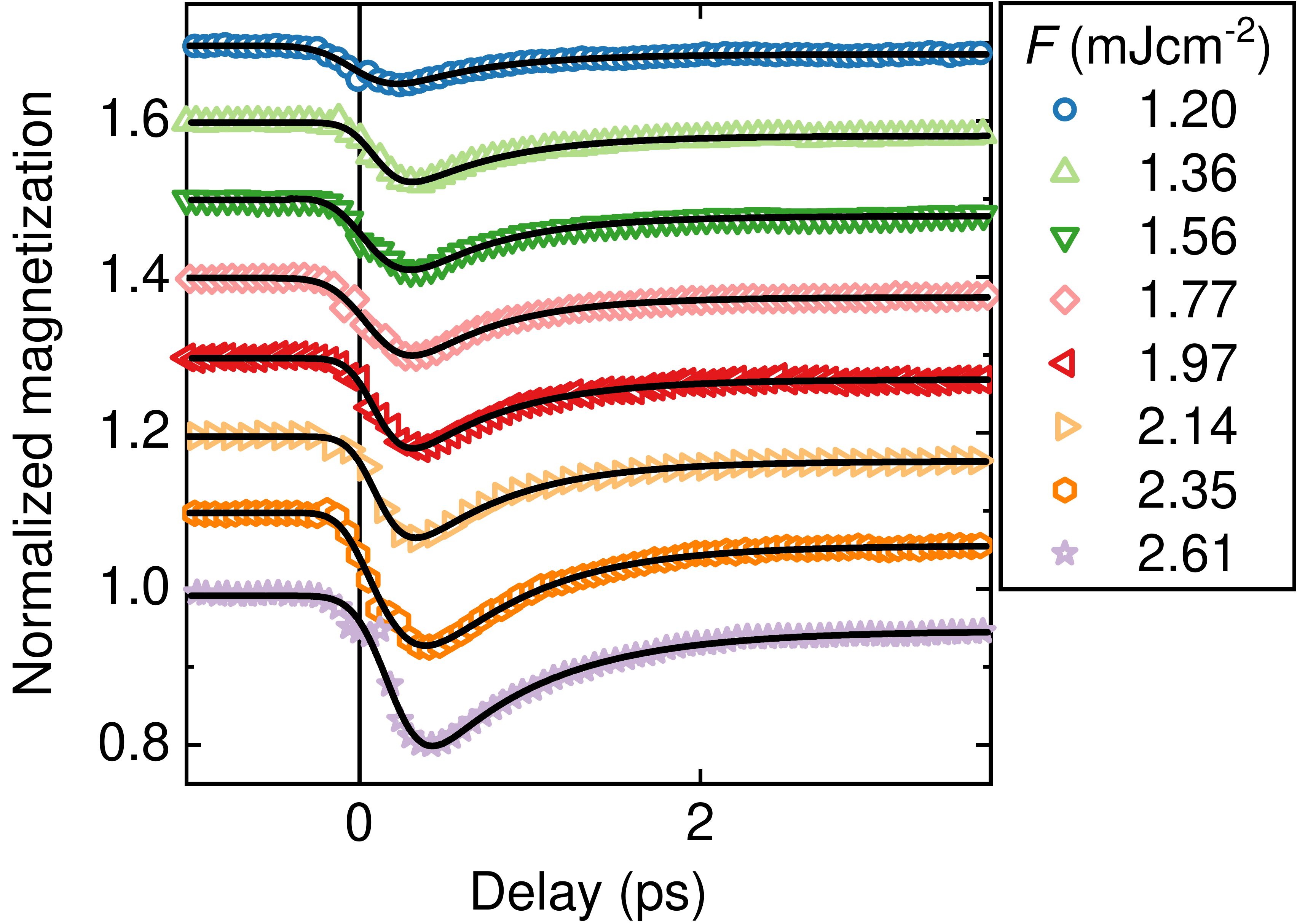}
\caption{\label{fig:S3}Complete dataset of the demagnetization measurements.}
\end{figure}

\begin{figure}[h!]
\includegraphics[width=12cm]{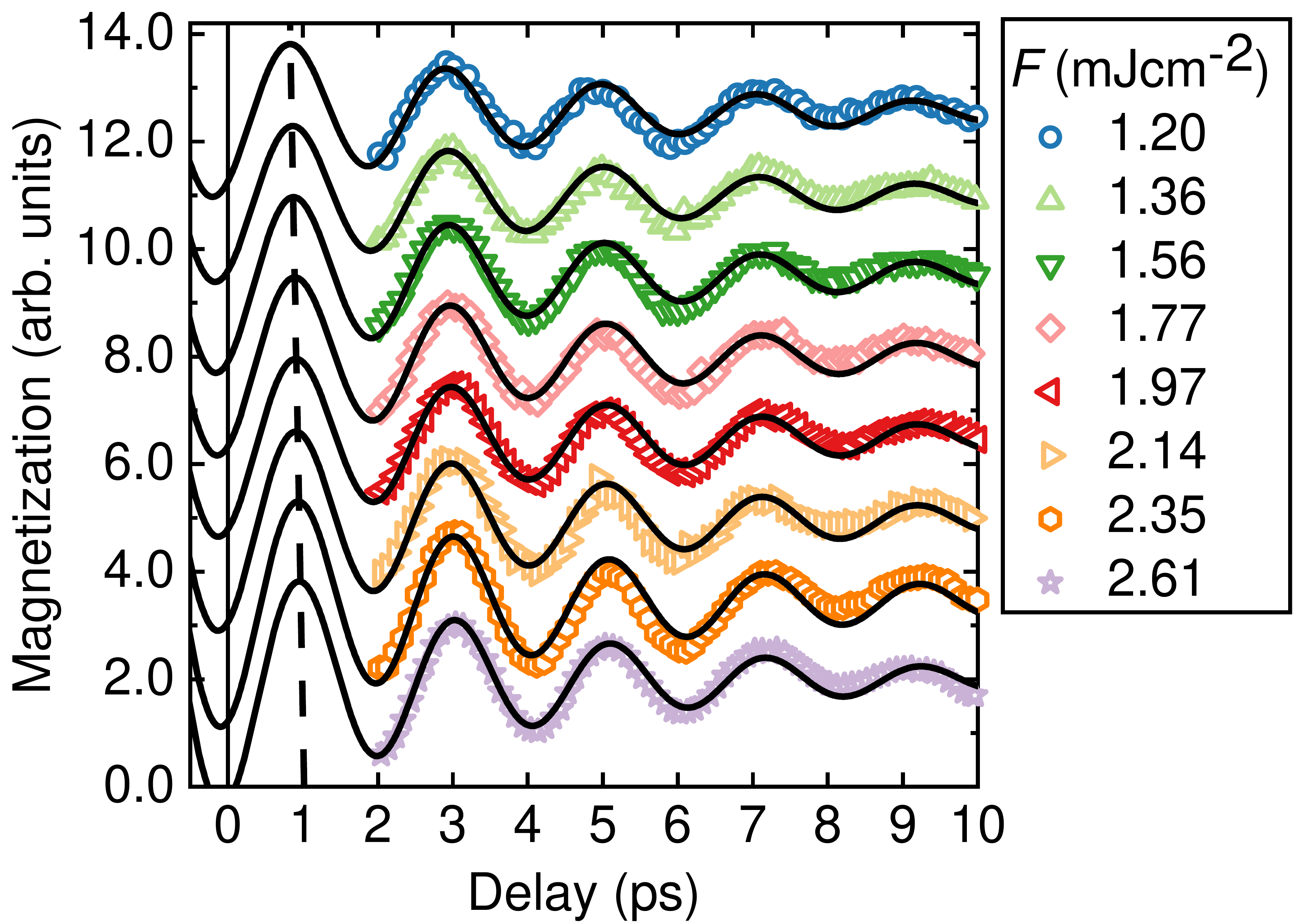}
\caption{\label{fig:S4}Complete dataset of the Thz spin wave measurements.}
\end{figure}

\begin{figure}[h!]
\includegraphics[width=12cm]{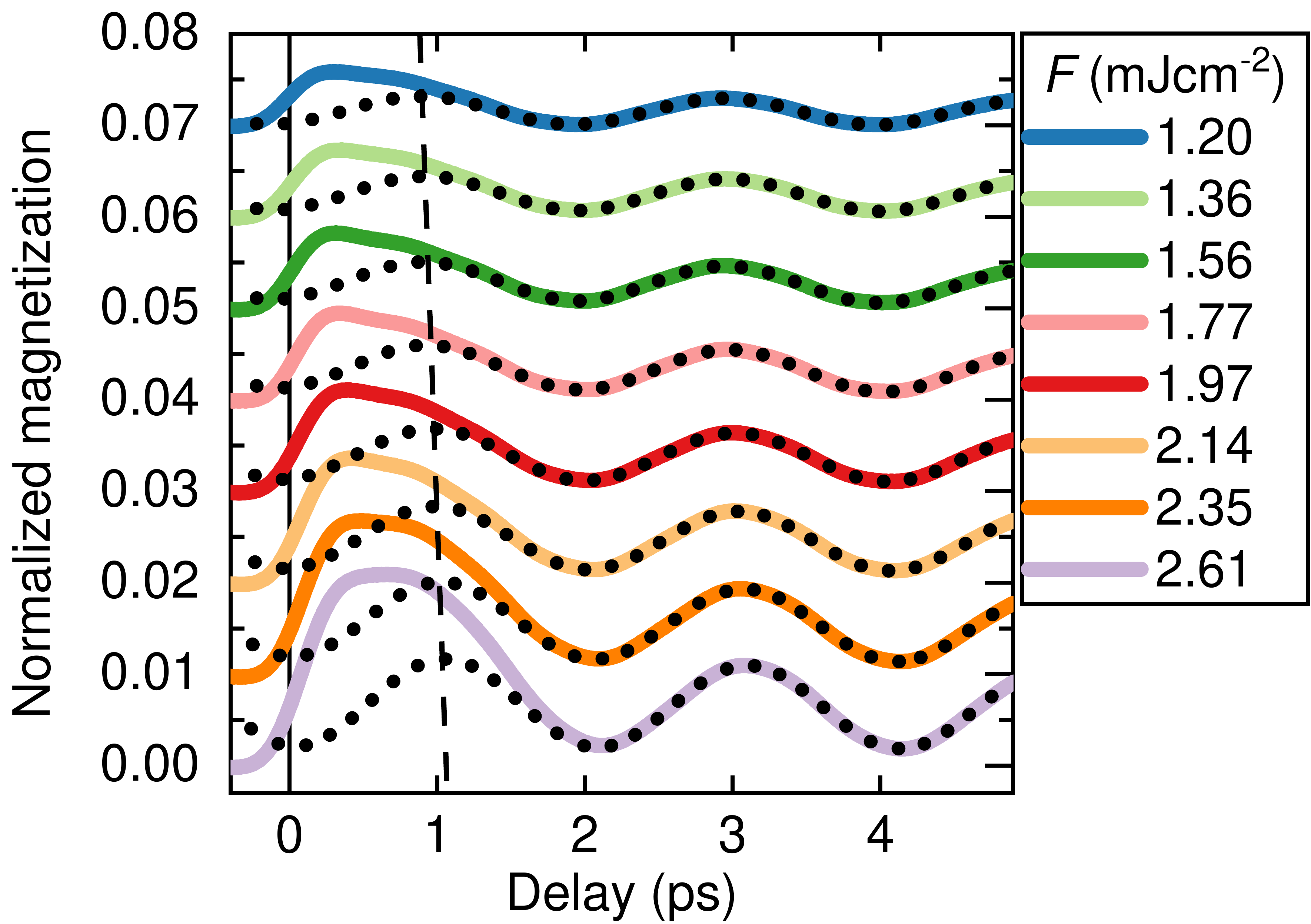}
\caption{\label{fig:S5}Complete dataset of the micromagnetic simulations.}
\end{figure}

\noindent For clarity, data for only four laser fluences was presented in the main text. In this section, the complete datasets used to construct fig.\,2e are shown. Fig.\,\ref{fig:S3}, \ref{fig:S4} and \ref{fig:S5} show the data corresponding to fig. 2a, 2b and 2d in the main text. 
\clearpage
\newpage
\section{Comparing Field-like and damping like torques}
\noindent In the MuMax$^3$ package, the spin transfer torque working on magnetization $\overrightarrow{\mathbf{m}}$ is defined as \cite{vansteenkiste2014design}
\begin{align}
\vec{\tau}_{\mathrm{SL}}&=\beta \frac{\epsilon-\alpha \epsilon^{\prime}}{1+\alpha^{2}}\left(\overrightarrow{\mathbf{m}} \times\left(\overrightarrow{\mathbf{m}}_{P} \times \overrightarrow{\mathbf{m}}\right)\right)-\beta \frac{\epsilon^{\prime}-\alpha \epsilon}{1+\alpha^{2}} \overrightarrow{\mathbf{m}} \times \overrightarrow{\mathbf{m}}_{P}\,, \\
\beta&=\frac{j_{z} \hbar}{M_{\mathrm{s}} e d}\,,\\
\epsilon&=\frac{P(\overrightarrow{\mathbf{r}}, t) \Lambda^{2}}{\left(\Lambda^{2}+1\right)+\left(\Lambda^{2}-1\right)\left(\overrightarrow{\mathbf{m}} \cdot \overrightarrow{\mathbf{m}}_{P}\right)}\,.
\end{align}
Here, $P$ is the fractional polarization of the spin current and $\overrightarrow{\mathbf{m}}_{P}$ the direction of the spin current polarization. In this work, $P=1$ and $\overrightarrow{\mathbf{m}}_{P}=\overrightarrow{\mathbf{z}}$ are used, given by the sample geometry. $\Lambda$ gives the angular dependence of $\epsilon$, which in turn represents the contribution of the damping like term. Here, we disregard this dependence for simplicity using $\Lambda=1$, which reduced $\epsilon$ to $P/2$. $\epsilon$' is the secondary spin-torque parameter, a measure for the contribution of the field-like term. The ratio between the prefactors in eq.\,S14 is directly proportional to the ratio between the two torque contributions, and is given by 
\begin{equation}
\frac{\vec{\tau}_{\mathrm{DL}}}{\vec{\tau}_{\mathrm{FL}}}=\frac{\epsilon-\alpha \epsilon^{\prime}}{\epsilon^{\prime}- \alpha \epsilon}\, ,
\end{equation}
where $\epsilon^{\prime}$ is used used as a free parameter in our simulations.
\newpage
\bibliography{Supp}